\documentclass[aps,pra,twocolumn,showpacs,nofootinbib]{revtex4-1}
\usepackage {amsmath}
\usepackage {amssymb}
\usepackage[dvips]{graphicx}
\usepackage{indentfirst}

\usepackage[utf8]{inputenc}

\usepackage[usenames]{color}
\usepackage{ulem}
\usepackage[colorlinks=true]{hyperref}

\newcommand{\rmi}{\mathrm{i}}
\newcommand{\rme}{\mathrm{e}}

\begin{document}

\title{Stability of out-of-phase solitons and laser pulse self-compression in active multi-core fibers}

\author{A.\,A.\,Balakin}
\author{A.\,G.\,Litvak}
\author{S.\,A.\,Skobelev}

\affiliation{Institute of Applied Physics of the Russian Academy of Sciences, 603950 Nizhniy Novgorod, Russia}

\date{\today}

\begin{abstract}
The out-of-phase soliton distribution of the wave field was found for a multicore fiber (MCF) from an even number of cores located in a ring. Its stability is proved both with respect to small wave field perturbations, including azimuthal ones and to small deformations of the MCF structure. As an example of using this soliton distribution, the problem of laser pulse compression in an active MCF is studied. The optimal fiber parameters, the minimum duration of the output pulse, and the compression length have been found, which are in good agreement with the results of numerical simulation. In order to achieve high energies in the output laser pulse, the requirements for MCF deformations are determined.
\end{abstract}

\maketitle
\section{Introduction}\label{sec:1}

Successful development of fiber-optic technologies in recent decades has stimulated the study of the possibility of replacing high-power solid-state lasers with equivalent laser systems based on fiber components, which can fundamentally change the attractiveness of relevant applied developments. This is due to small sizes of fiber laser system, the ease of controlling them, and reliability and stability of their operation. In particular, the use of an array of independent active light-guides is proposed as a promising method for producing laser pulses with an extremely high power level \cite{Tajima14,Tajima13}. The maximum achievable radiation power in each fiber is limited, but the total power can be arbitrarily large in the case of coherent summation of pulses from many fibers. One of the difficulties of this approach is high sensitivity of the method of coherent field summation to various disturbing factors. It is required to maintain a constant phase difference between laser channels under conditions of random variations in the radiation phase in each channel. Recent work \cite{core_8_activ,core_64_low,core} demonstrated experimentally the possibility of synchronizing laser radiation at the output of a number of independent light-guides. At this, only 8 light-guides can be synchronized for intense wave packets.

Successful development of technologies for manufacturing multi-core fibers (MCF), which consist of identical weakly coupled optical cores located equidistantly, stimulates studies focused on the possibility of coherent propagation of laser radiation with a total power noticeably greater than that transmittable in a single-core optical fiber. Using MCF allows one to split the total high power into cores with a power below any unwanted non-linear effects that result in fiber damage. In other words, laser beams in each core can be safely transported below the threshold of harmful nonlinear effects, while the total coherent power can be very high. This stimulated the study of nonlinear wave processes in spatially periodic media being sets of a large number of weakly coupled optical cores.

Unfortunately, these expectations have failed. As shown by theoretical and experimental studies \cite{Cheskis03,Christodoulides88,Balakin16,	Eilenberger2011,Turitsyn15,Minardi,Kivshar94}, such systems have its own critical power/energy, at which self-focusing of the quasi-homogeneous distribution of the discrete wave field occurs, and the field disintegrates into a set of incoherent structures~\cite{Balakin18}. Despite that, a number of interesting results were obtained in this direction related to studying the possibilities of generating a supercontinuum \cite{Tran,Panagiotopoulos}, reducing the duration of laser pulses \cite{Aceves,Turitsyn16,Rubenchik,Balakin18,	Tran14,Cheskis,Eisenberg2001,Skobelev18,Rubenchik2013}, controlling the structure of the wave field \cite{Turitsyn12,Skobelev2018,Boris,Lederer2008,Christodoulides2003}, and producing light bullets \cite{Eilenberger2011,Minardi,Turitsyn15,Mihalache,Kartashov}. The presence of discrete critical power/energy leads to the fact that the power/energy of the used laser radiation is also limited by the destruction threshold or self-focusing in a separate core.

\begin{figure}[tb]
	\includegraphics[width = 0.9\linewidth]{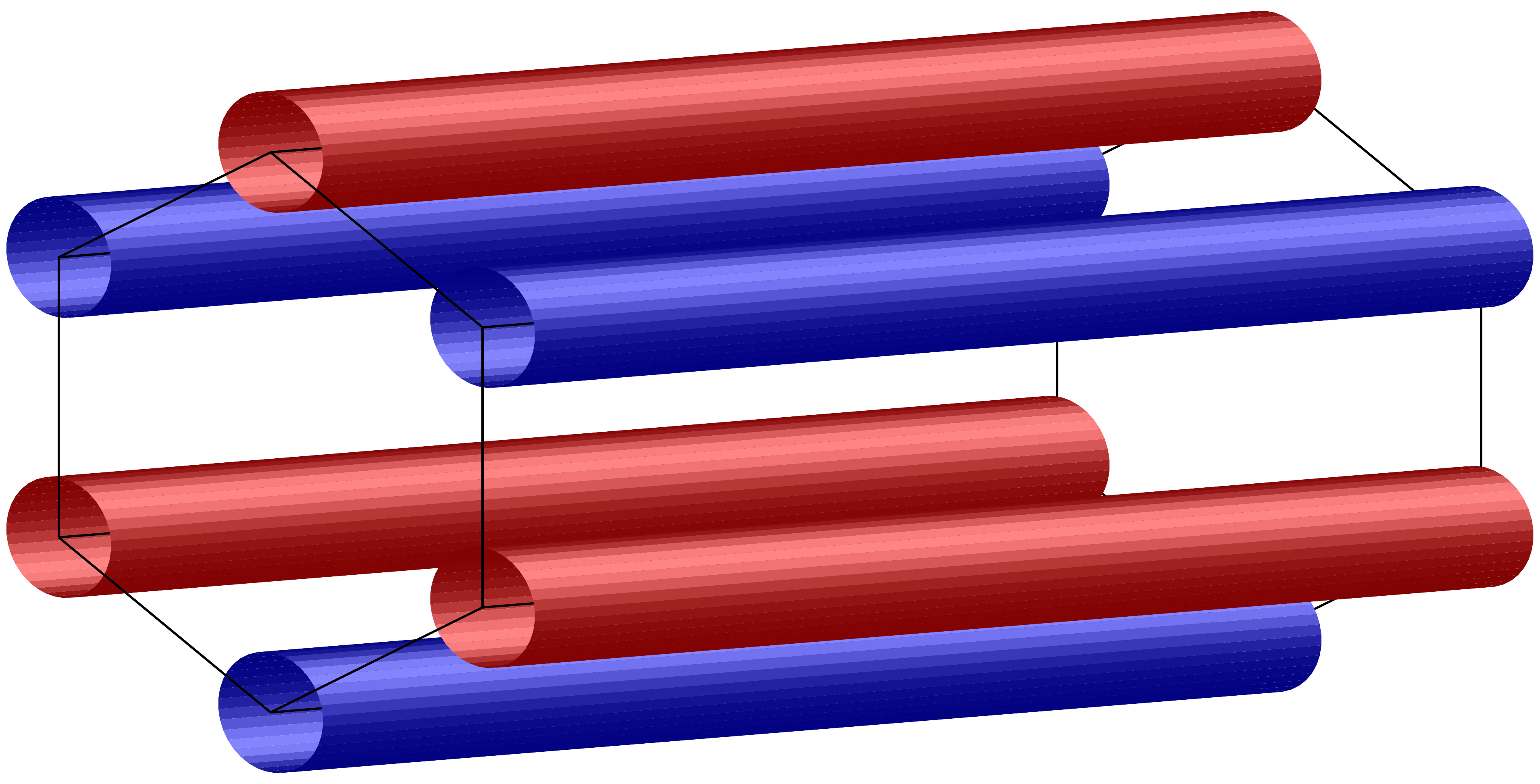}
	\caption{(Color online) Schematic illustration of the considered MCF with cores arranged in a rind. Color reflects the out-of-phase distribution of the wave field $u_n \propto (-1)^n$.}\label{ris:ris1}
\end{figure}

Recently, the research focus has shifted to MCF with a small number of cores, e.g., MCF consisting of a central core and an even number ($2N$) of cores located in a ring \cite{Turitsyn16,Turitsyn12,Skobelev2018,Rubenchik2013,Rubenchik} (see Fig.~\ref{ris:ris1}). In such MCFs, stable inhomogeneous stationary nonlinear wave field distributions with a total power much larger than the critical self-focusing power were found \cite{Skobelev2018}. This includes both the well-known solution localized in one core and solutions, which use all cores ($\pm$-mode, when the phase differs by $\pi$ in neighboring cores), a half of them ("<crown">-like distribution, which has the form of in-phase maxima through the core), or only several optical cores (mirror-symmetric distributions). In the case of the $\pm$-mode, the total radiation power can be many ($2N$) times higher than the critical self-focusing power in a continuous medium.

These successful efforts motivate the search for stable spatio-temporal soliton solutions for the purpose of coherent propagation of unchanged laser pulses in all available cores of such MCF. The total energy of such a nonlinear structure can significantly exceed the energy of the NSE soliton in a single core. The existence of these nonlinear solutions allows us to easily generalize the well-known methods of laser pulse compression in a single core to the case of MCF, i.e., to take a significant step in solving the problem of formation of high-energy (sub-$\mu$J), short-duration laser pulses in systems based entirely on the fiber design. Note that in the case of a single-core fiber in the region of anomalous group velocity dispersion (in the region of existence of NSE solitons), many schemes for decreasing the wave packet duration are well developed: (i) use of a high-order soliton \cite{Mollenauer,Li10,Agrawal}; (ii) adiabatic change in the parameters of a soliton-like laser pulse in a waveguide system with monotonically decreasing dispersion \cite{Chernikov92,Agrawal}, (iii) in a medium with amplification \cite{Lisak,Nakazawa}, and others.

The goal of this work is to study the existence and stability of the spatio-temporal soliton in a MCF being a ring of $2N$ identical weakly coupled cores (see Fig.~\ref{ris:ris1}), with the aim of the possibility of coherent propagation of unchanged laser pulses in all available MCF cores. As an example of the use of this soliton, we analyze the adiabatic decrease in the soliton duration in the active MCF with a finite Gaussian amplification band under the conditions of a non-stationary nonlinear response of the medium. Based on the variational approach, the minimum duration and, accordingly, the maximum energy of the laser pulse were determined. The possibility of generating a laser pulse of 55~fs duration with a sub-$\mu$J level of energy at the MCF output at a length of 7~m at a wavelength of 2~$\mu$m doped with thulium was demonstrated. The requirements for the MCF strains and the spread of the gain in the cores for achievement high energies in the output laser pulse are determined.

The work plan is as follows. In Section~\ref{sec:2}, the basic equations are formulated. In Section~\ref{sec:4}, the spatiotemporal soliton in the passive MCF is found, and its stability with respect to azimuthal perturbations is analyzed. In Section~\ref{sec:3}, we consider the stability of the solution found with respect to small deformations of the MCF structure and the spread of the gain across the cores. In Section~\ref{sec:5}, we perform a qualitative analysis of the adiabatic decrease in the duration of the found spatio-temporal soliton in the active MCF with a finite Gaussian gain band allowing for the nonstationarity of the nonlinear response. Section~\ref{sec:6} presents the results of a numerical simulation of the laser pulse self-compression in a silica MCF made of six cores doped with thullium. In the Conclusion, the main results of the work are formulated.

\section{Formulation of the problem}\label{sec:2}

Let us consider the self-action of sub-picosecond laser pulses in an active multi-core fiber (MCF) with a finite gain bandwidth taking into account the instantaneous electron (Kerr) and delayed molecular (Raman) nonlinearities. Next, we will consider MCF made of an even ($2N$) number of cores arranged in a ring. Figure~\ref{ris:ris1} shows schematically an MCF fiber with $N = 3$. The analysis will be based on the standard theoretical model \cite{Aceves,Balakin18,Christodoulides88,Lederer2008,Balakin16,Kivshar,Turitsyn16}, in the framework of which it is assumed that the fundamental guided modes of the optical cores oriented parallel to the $z$-axis are weakly coupled. In this case, the propagation of laser pulses in the MCF can be described approximately as a superposition of fundamental modes localized in each core
\begin{equation}\label{eq:1}
	E \simeq \sum\limits_{n}{E}_n(z,t)\phi(x-x_n,y-y_n)e^{\rmi k_n z-\rmi\omega t }+c.c. ,
\end{equation}
where $\phi(x,y)$ is the structure of the lowest spatial mode in the core, ${E}_n$ is the envelope of the electric field strength in the $n$-th core, which slowly changes along the $z$-axis. We assume that the central frequency of the laser pulse lies in the region of the anomalous group velocity dispersion. We neglect the gain saturation of the wave packet, since fiber laser systems have a high saturation energy. For example, the saturation energy of erbium fiber laser systems is approximately 10~$\mu$J \cite{Agrawal90,Agrawal_1996}, and the energy of the considered laser pulses is less than 10~nJ per the core.

The evolution of the wave field envelope in the $n$-th core can be affected by linear dispersion and the Kerr nonlinearity of a single core, amplification in the active medium, and interaction with the nearest neighboring cores that arises due to weak overlapping of their fundamental modes. Assuming that the cores are weakly coupled, we obtain the following system of equations for the envelope of the electric field ${E}_n$ in the $n$-th core \cite{Agrawal_1996,Turitsyn16,Balakin18,Skobelev18}:
\begin{multline}\label{eq:2}
	\rmi\dfrac{\partial{E}_n}{\partial z}+\rmi \dfrac{\partial k_n}{\partial \omega}\dfrac{\partial{E}_n}{\partial t}=\dfrac{\beta_n}2\dfrac{\partial^2{E}_n}{\partial t^2}+\sum\limits_{m=1}^{2N}\chi_{mn}{E}_m+\\+\Gamma_n P_n^\text{NL}+ \dfrac{\rmi\omega_0}{4\pi n_{0n} c}\int_{-\infty}^{+\infty}X_n(\omega){E}_n(z,\omega)\rme^{-\rmi\omega\tau}d\omega .
\end{multline}
Here, the subscript $n$ varies from $1$ to $2N$, $\omega_0$ is the carrier frequency of the laser pulse, $c$ is the speed of light, the coefficient $\chi_{mn}=\chi_{nm}$ determines the coupling between the cores $m$ and $n$, $\chi_{nn} = k_n$ is the wave number in the cores, $n_{0n}$ and $\Gamma_n$ are the linear and nonlinear refractive index in the $n$-th core, respectively, $X_n(\omega)$ is the imaginary part of the susceptibility in the $n$-th core, and $\beta_n = \partial^2 k_n / \partial \omega^2$ is the group velocity dispersion in the $n$-th core. The term $P_n^\text{NL}$ takes into account the time dependence of the nonlinear response of the fiber \cite{Kivshar,Agrawal}
\begin{equation}\label{eq:2_1}
	P_n^\text{NL}=(1-f_R)|{E}_n|^2{E}_n+f_R{E}_n\int_0^\infty|{E}_n(t-t')|^2h_R(t')dt' ,
\end{equation}
where $f_R$ represents the partial contribution of the SRS inertial response to nonlinear polarization. The Raman response function $h_R$ is responsible for the Raman gain and can be determined from the experimentally measured Raman spectrum in silica fibers. The approximate analytical form of this function is as follows \cite{Kivshar,Agrawal}:
\begin{equation}
	h_R(t)=\dfrac{\tau_1^2+\tau_2^2}{\tau_1\tau_2^2}\exp(-t/\tau_2)\sin(t/\tau_1) .
\end{equation}
For silica fiber, we have $\tau_1 = 12.2$~fs, $\tau_2 = 32$~fs, and $f_R = 0.18$ \cite{Kivshar}. When analyzing the nonlinear dynamics of laser pulses with a duration of $\tau_p \gg \tau_2$, expression~\eqref{eq:2_1} can be simplified:
\begin{equation}\label{eq:2_2}
	P_n^\text{NL}\approx |{E}_n|^2{E}_n-\tau_R{E}_n\dfrac{\partial |{E}_n|^2}{\partial t} , \qquad \tau_R = \dfrac{2f_R\tau_1^2\tau_2}{\tau_1^2+\tau_2^2}.
\end{equation}

We assume that all cores are the same, i.e. nonlinearity coefficients $\Gamma_n \equiv \Gamma$, amplification $X = X_n$ and coupling are the same for all cores $\chi_{nm}=\widehat{\chi} \delta_{n,m\pm1}$. In the accompanying coordinate system moving with the group velocity of the wave packet, the system of equations \eqref{eq:2} taking into account the expression \eqref{eq:2_2} can be written in dimensionless variables as follows:
\begin{multline}\label{eq:3}
	\rmi \dfrac{\partial u_n}{\partial z} = \dfrac{\partial^2 u_n}{\partial\tau^2}+|u_n|^2 u_n-\mu u_n\dfrac{\partial |u_n|^2}{\partial\tau}+\chi(u_{n+1}+u_{n-1})+\\
		+\delta h_n u_n+\dfrac{\rmi}{2\pi}\int_{-\infty}^{+\infty}G(\omega)\, S_n(\omega) \rme^{-\rmi\omega\tau}d\omega.
\end{multline}
Here, the longitudinal coordinate $\tau = t - \frac{\partial k}{\partial \omega} z$ and the evolutionary coordinate $z$ are normalized to the characteristic laser pulse duration $\tau_\text{in}$ and the corresponding dispersion length $z_0 = \frac{2\tau_\text{in}^2} {\partial^2 k/\partial \omega^2}$, $u_n =\rme^{-\rmi h_n z} {E}_n \sqrt{\Gamma z_0}$ is the complex amplitude of the envelope of the wave packet in the $n$-th core, $h_n=k_n^{(0)} z_0$, $k_n=k_n^{(0)}+\delta k_n$, $\chi=\widehat{\chi}z_0$, $\delta h_n=\delta k_n z_0$; $S_n = \int_{-\infty}^{+\infty} u_n(\tau)\rme^{i\omega\tau}d\tau$ is the envelope spectrum; $\mu = \tau_R/\tau_\text{in}$, $G = \frac{\omega_0 z_0}{2 n_0 c} X(\omega)$. The Gaussian distribution will be chosen as the gain profile of the active medium $G(\omega)$
\begin{equation}\label{eq:4}
	G(\omega)=\gamma\exp\left(-{\omega^2}/{\Omega^2}\right) .
\end{equation}

The second term in Equation \eqref{eq:3} describes the linear dispersion of the medium, the third term is responsible for the Kerr nonlinearity, and the fourth term is the inertia of the nonlinear response. Further terms are responsible for coupling with neighboring cores. The last term describing the process of amplification of a laser pulse in an active medium, is represented through the spectral gain $G(\omega) $. In the case of a single-core fiber, the reduced system of equations corresponds to the well-known equation from \cite{Agrawal_1996}.

The applicability of Eqs.~\eqref{eq:3} is limited by the approximation of the single-mode wave field propagation in each core. It is violated when the radiation power in any of the cores $\mathcal{P}_n=|{E}_n|^2\iint \phi^2 dxdy$ becomes close to the critical power of self-focusing in a homogeneous medium $\mathcal{P}_\text{cr}$.

\section{Spatio-temporal soliton}\label{sec:4}

A number of stable nonlinear solutions for wave beams propagating in the considered MCFs were found in \cite{Skobelev2018}. The most interesting of these is the $\pm$-mode $u_n \propto (-1)^n$, which provides transportation of maximum power at a given field amplitude. Moreover, this solution is stable and exists at all amplitudes.

The generalization of the $\pm$-mode to the pulsed case is simple, since wave field amplitudes in all cores are the same. Indeed, equations \eqref{eq:3} in the absence of amplification $G = 0$ and an unsteady nonlinear response $\mu = 0$ have a solution in the form
\begin{equation}\label{eq:6}
u_n(z,\tau)=(-1)^n\dfrac{\sqrt{2}b\rme^{\rmi(2\chi-b^2) z}}{\cosh(b\tau)}.
\end{equation}

To prove the stability of the manifold \eqref{eq:6}, we use the {\it second Lyapunov method}. It says that the variety of solutions will be stable if it is possible to find the Lyapunov functional $\mathcal{F}[u]$ that satisfies the following requirements: (i) it is always positive $\mathcal{F}[u] \ge 0$; (ii) the derivative is less than or equal to zero $\frac{d}{dz} \mathcal {F} \le 0$; and (iii) $\mathcal {F} [u_s] = 0$ \emph{only} on this manifold.

The key point here is the requirement of uniqueness, $\mathcal{F}[u]=0$, for the analyzed solution only. For example, for the nonlinear oscillator $\ddot{x}-x+x^3=0$, the Lyapunov functional $\mathcal{F} = (\dot{x}^2-x^2 + x^4/2 - \mathcal{H}_0)^2$ shows the stability of the manifold of nonlinear oscillations at $\mathcal{H}_0 \ne 0$ with different initial phases. However, in the case of $\mathcal{H}_0=0$, it only proves the stability of \emph{all} separatrices as a whole. Moreover, each individual separatrix (left or right) is unstable. This reflects the presence of a stochastic layer in the small vicinity of the saddle point and separatrices.

Equations~\eqref{eq:3} with $\mu = G = 0$ preserve the Hamiltonian $\mathcal{H}$ and the energy $W$:
\begin{subequations}\label{eq:7}
	\begin{gather}
	\mathcal{H} = \sum_{n=1}^{2N} \int\limits_{-\infty}^{+\infty} \left[\left|\dfrac{\partial u_n}{\partial\tau}\right|^2 - \dfrac{|u_n|^4}{2} - \chi (u_n u_{n+1}^* + u_{n+1} u_n^*) \right] d\tau,\label{eq:7a} \\
	W = \sum_{n=1}^{2N} \int\limits_{-\infty}^{+\infty} |u_n|^2 d\tau . \label{eq:7b}
	\end{gather}
\end{subequations}
The Hamiltonian $\mathcal{H}$ is well suited for the role of the Lyapunov functional, but it is not formally bounded below. Indeed, an increase in the field amplitude due to a perturbation can lead to an increase in the nonlinear term and a decrease in the value of $\mathcal{H}$. Let us find the minimum of the Hamiltonian $\mathcal{H}$ in the class of functions that conserve energy using the method of indefinite Lagrange multipliers. The first variation of the functional $\mathcal{R}[u] = \mathcal{H}[u] + \lambda W[u]$ with respect to $u_n^*$ gives the equation
\begin{equation}
-\dfrac{\partial^2u_n}{\partial\tau^2} - |u_n|^2 u_n - \chi (u_{n+1} -2 u_n + u_{n-1}) + \lambda u_n = 0 
\end{equation}
for the soliton manifold ($\lambda_m = \lambda - 2 \chi \cos \kappa_m$)
\begin{equation}\label{eq:8}
u_n^\text{(m)} = \rme^{\rmi \kappa_m n} \dfrac{\sqrt{2\lambda_m}}{\cosh \sqrt{\lambda_m} \tau}, \quad W_m = 8N\sqrt{\lambda_m}.
\end{equation}
The Hamiltonian value for this solution is
\begin{multline}\label{eq:H^sol}
\mathcal{H}^\text{sol}_m = -16N \chi \sqrt{\lambda_m} \cos \kappa_m - \dfrac{8N}{3}\lambda_m^{3/2} =\\=
-2\chi W_m \cos \kappa_m-\dfrac{W_m^3}{192 N^2} .
\end{multline}

Then, the functional
\begin{equation}\label{eq:F_m}
\mathcal{F}_m[u] = \left(\mathcal{H}[u]+\dfrac{W[u]^3}{192 N^2} + 2 \chi W[u] \cos \kappa_m \right)^2
\end{equation}
satisfies the conditions of $\mathcal{F}_m[u_n^\text{sol}] \ge 0$ and $d\mathcal{F}_m[u]/dz=0$. The main difficulties arise with proving the uniqueness of the functional turning to zero only on solution \eqref{eq:8} and nowhere else.

\begin{figure}
	\includegraphics[width = \linewidth]{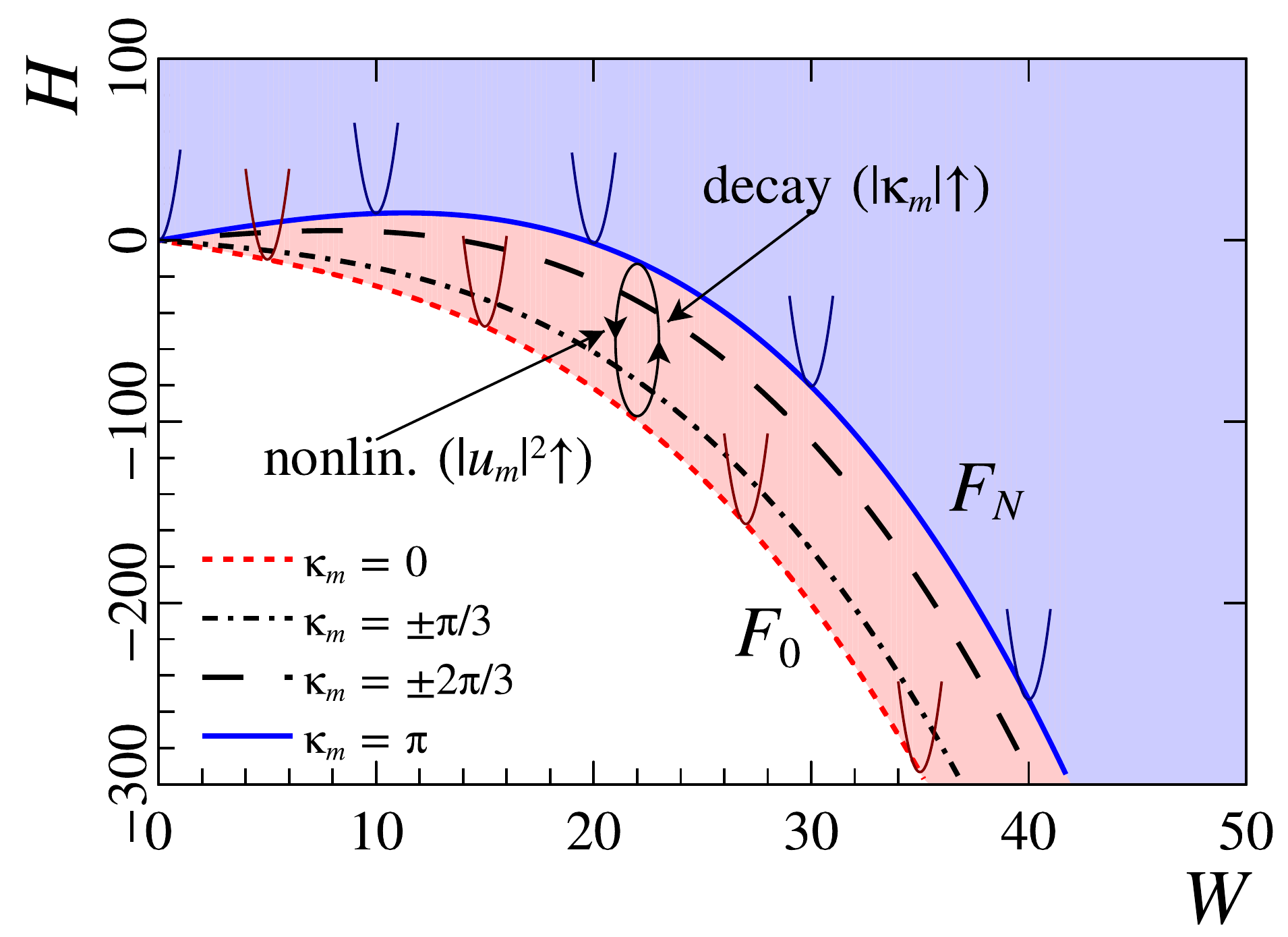}
	\caption{(Color online) Dependence \eqref{eq:7} of the Hamiltonian $\mathcal{H}$ on the energy $W$. The curves correspond to the extremums \eqref{eq:H^sol} for various $m$. The arrows show qualitative explanation of instability at $m=0$.}\label{ris:fig_Lyapunov}
\end{figure}

We first consider the simplest case of $N=1$. In this case, there are only two solutions with $\kappa_m = \pi$ (out-of-phase) and $\kappa_m = 0$ (in-phase). Figure~\ref{ris:fig_Lyapunov} shows the dependence of $\mathcal{H}_m^\text{sol}$ \eqref{eq:H^sol} on $W$, which corresponds to the minimum value of the Hamiltonian $\mathcal{H}$ at a fixed energy value $W$. Here, the blue solid curve corresponds to the case $\kappa = \pi$, and the red dashed line corresponds to $\kappa = 0$. On these curves, the functional $\mathcal {F}$ vanishes. Recall that the quantities $\mathcal{H}$ and $W$ are integrals of the problem. Note that the value of the Hamiltonian $\mathcal{H}_N^\text{sol}$ in the solution at $\kappa_m = \pi$ is always greater than the value $\mathcal {H}_0^\text{sol}$ at $\kappa_m = 0$ (see Fig.~\ref{ris:fig_Lyapunov}). If we add a small perturbation of the field to the solution $\tilde{u}=u^\text{(m)}+\delta u$, then the resulting wave field can turn into a soliton with different parameters ($\lambda, \kappa_m$) and some small ripples (the variety of solutions is stable), or will fall apart completely (unstable). Note that the perturbed field corresponds to new values of the energy $W[\tilde{u}]$ and the Hamiltonian $\mathcal{H}[\tilde{u}]$.

An increase in the intensity difference in the cores for case of the solution at $\kappa_m = 0$ leads to an increase in the nonlinear term and thereby to a decrease in the Hamiltonian $\mathcal{H}$ \eqref{eq:7a}. This change in the Hamiltonian is easily compensated by a decrease in the term that contains $\chi$ (for example, the appearance of a phase difference due to the decay to $\kappa_m = \pi$). As a result, the same values of $\mathcal{H}$ and $W$ (that is, $\mathcal {F}_0 = 0$) correspond to different distributions. Thus, the in-phase solution is unstable and tends to beats with radiation capture in one core. Contrary, the value of the Hamiltonian in the solution at $\kappa_m = \pi$ is always greater than its value at $\kappa_m = 0$. This means that a distribution with $\kappa_m = \pi$ cannot decay into a distribution with smaller $\kappa_m = 0$. Therefore, an increase in the nonlinear term at $\kappa_m = \pi$ cannot be compensated and leads to $\mathcal{F}_N>0$. In other words, the values of $\mathcal{H} \ne \mathcal{H}_N^\text{sol}$ and $W^\text{sol}$ correspond to different distributions. Thus, the out-of-phase soliton distribution is stable.

Now, we get back to the case of $N \geq 2$. An increase in the number of cores leads to appearance of additional modes at $\kappa_m = m \pi/N$. Again, it is possible to show the uniqueness of the functional  $\mathcal{F}$ \eqref{eq:F_m} turning to zero only for the $\pm$-mode with $\kappa_m = \pi$ (that is, $m = N$). For all other $m \ne N$, functional \eqref{eq:F_m} can be set to zero for $u_n \ne u_n^{(m)}$. This fact reflects the presence of filamentation instability at $\kappa_m \ne \pi$. The situation in this case is similar to the classical filamentation instability, which develops on a quasi-homogeneous wave packet with constant energy and Hamiltonian integrals.

Indeed, let us investigate the obtained solution with respect to filamentation instability. For $N \geq 2 $ and a wave field with perturbations having the form
\begin{equation*}\label{eq:9}
u_n^{(m)}=\left[f_m \rme^{\rmi\kappa_m n}+\delta_s\rme^{\rmi \lambda z+\rmi \kappa_s n} \right]\rme^{-\rmi\left(2\chi \cos\kappa_m+f_m^2 \right)z} 
\end{equation*}
Assuming $|\delta_s|\ll f_m$, we get real eigenvalues
\begin{multline}\label{eq:10}
\lambda^2=4\chi\left[\cos\kappa_m-\cos\kappa_s \right]\times\\\times \left[\chi(\cos\kappa_m-\cos\kappa_s)- f_m^2\right].
\end{multline}
For $\kappa_m \neq \pi$, a stable coherent radiation propagation regime ($\lambda^2>0$) is realized only when the wave field amplitude in the core is less than the critical value $f_m <f_\text{cr}$. For example, in the case of an isotropic field distribution on the ring ($\kappa_m = 0$), the filamentation is absent only at amplitudes
\begin{equation}\label{eq:11}
f_0<f_\text{cr}=\sqrt{2\chi}\sin\dfrac{\pi}{2N}\mathop{\approx}\limits_{N \gg 1} \frac{\pi}{N} \sqrt{\frac\chi2}.
\end{equation}

\begin{figure}
	\includegraphics[width = \linewidth]{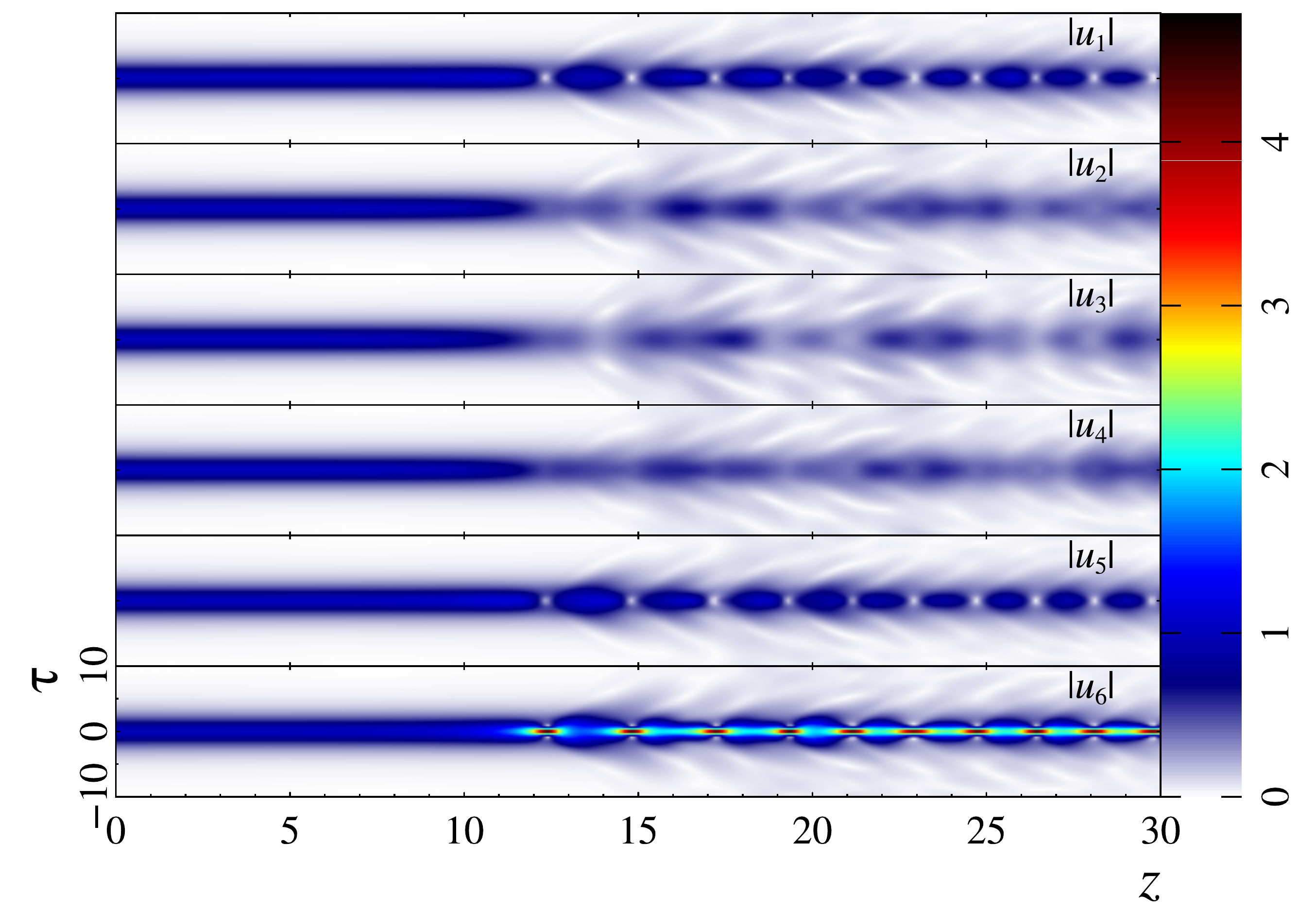}
	\caption{(Color online) Dynamics of the wave field envelope $|u_n|$ in a 6-core MCF ($N = 3$) with the coupling coefficient $\chi = 1$. The initial distribution is $ u_n = \sqrt{2} b/\cosh (b \tau)$ with $b = 0.7$.}\label{ris:ris2}
\end{figure}

Next, we turn to the results of numerical simulations in order to verify the above qualitative analysis of the stability of solution \eqref{eq:8} for $\kappa_m \neq \pi$. Figure~\ref{ris:ris2} shows the evolution dynamics of a laser pulse with initial distribution \eqref{eq:8} in a 6-core MCF ($N=3$) with the coupling coefficient $\chi = 1$ in the case of an isotropic field distribution on the ring ($\kappa_m = 0$) with the amplitude $b = 0.7$. With these parameters, the initial amplitude is greater than the critical value ($\sqrt{2} b> f_\text{cr} = 1/\sqrt{2}$). The results of numerical simulations confirm the instability of solution \eqref{eq:8} at $\kappa_m = 0$. As the wave packet propagates in the medium, the radiation is captured in an arbitrary core (at $z \sim 12$). This leads to a decrease in the duration of the laser pulse by several times \cite{Rubenchik,Turitsyn16}.

\section{Resistance to MCF deformation}\label{sec:3}

\begin{figure*}
	\includegraphics[width = \linewidth]{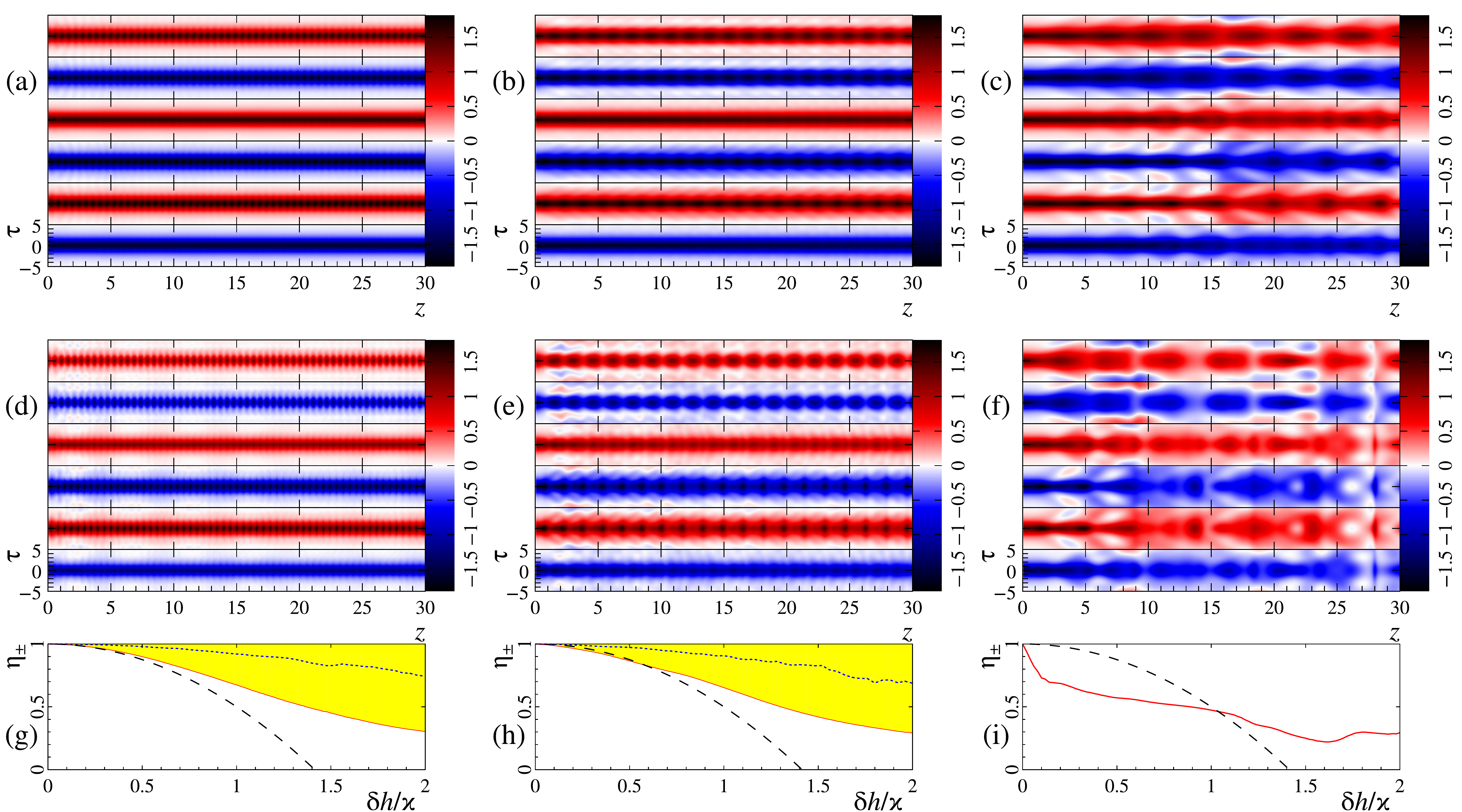}
	\caption{(Color online) The envelope dynamics of the wave field $Re[u_n(z,\tau)]$ in 6-cores MCF ($N = 3$) in the case of fiber bending $h_n=h+\delta h\sin(\pi n/N)$ for different values of the coupling coefficient $\chi$: {\bf (a, d, g)} $\chi = 10$; {\bf (b, e, h)} $\chi = 3$; {\bf (c, f, i)} $\chi = 1$. The top row corresponds to $\delta h/\chi = 0.1$, the middle row, to $\delta h/\chi = 0.3$. The space-time soliton $u_n = (-1)^n \sqrt{2} b/\cosh(b \tau)$ was injected at the MCF input with $b = 1$. {\bf (g, h, i)} The dependence of the fraction of energy $\eta_\pm$ in the $\pm$-mode on the value of the parameter $\delta h/\chi$. Here, the black dashed line is estimate \eqref{eq:pm_dP_lin} found for the beam problem. The red line shows the results of numerical modeling. The yellow region is the fraction of energy in the $\pm$-mode, when $\delta h_n$ is random and lies in the range $0 \leq \delta h_n \leq \delta h$. Blue dots represent the fraction of energy in the $\pm$-mode after averaging over the ensemble.} \label{ris:soliton_pm_chi}
\end{figure*}

The stability of the found spatio-temporal soliton \eqref{eq:6} with respect to small perturbations of the wave field suggests its stability with respect to small deformations of the MCF structure ($\delta h_n \neq 0$). The results of numerical modeling confirm this conclusion. As an example, consider two cases: (i) the linear gradient $\delta h_n = \delta h \sin (\pi n/N)$; (ii) the value of $\delta h_n$ is random and lies in the interval $0 \leq \delta h_n \leq \delta h $.

Figure \ref{ris:soliton_pm_chi} shows the results of the numerical simulation of the evolution of wave packet \eqref{eq:6} in a 6-core MCF ($N = 3$) for different values of the coupling coefficient $\chi$: {\bf (a,d)} $\chi=10$, {\bf (b,e)} $\chi=3$, and {\bf (c,f)} $\chi=1$. Laser pulse \eqref{eq:6} with $b=1$ was injected at the input of MCF. The top row corresponds to the case $\delta h/\chi = 0.1$, the middle row, to $\delta h/\chi = 0.3$. The gradient of the refractive index ($\delta h_n = \delta h \sin (\pi n / N)$) leads to the fact that the field distribution becomes inhomogeneous along the MCF ring. The field amplitude has maximal values in the cores, where $\delta h_n = -\delta h$ and, accordingly, the minimal values at $\delta h_n = \delta h$. In Fig.~\ref{ris:soliton_pm_chi} {\bf (g, h, i)}, the red line shows the dependence of the fraction of energy $\eta_\pm = 1/(1+\delta W/W)$ in the $\pm$-mode on the modulation depth $\delta h/\chi$ for different values of the coupling coefficient $\chi$, where $\delta W/W$ is the fraction of energy in other modes ($\kappa_m \neq \pi$). It can be seen that with an increasing depth of the modulation $\delta h / \chi$, the fraction of the energy in the $\pm$-mode decreases obeying a parabolic law.

Figure~\ref{ris:soliton_pm_chi}{\bf (a, b)} shows that deformations of the MCF have nearly no affects on the dynamics of soliton propagation in the case of $\chi \geq 3$ and the modulation depth $\delta h/\chi = 0.1$. Over the entire interval of the numerical calculation, the soliton duration does not change ($z \ggg z_{dis}$, where $z_{dis} \simeq 1$ is the dispersion length) and the $\pm$-mode is not disturbed (the relative phase difference in neighboring cores is $\pi$). The fraction of the energy in the $\pm$-mode is $\eta_\pm \approx 0.98$. With an increasing modulation depth (see Fig.~\ref{ris:soliton_pm_chi}{\bf (d, e, f)}) of the refractive index ($\delta h/\chi = 0.3$), the temporal structure remains unchanged only at a large coupling coefficient $\chi = 10$ (see Fig.~\ref{ris:soliton_pm_chi}{\bf (d)}). Only some beats in the amplitude with respect to $z$ are observed. The fraction of the energy in the $\pm$-mode is $\eta_\pm \approx 0.82$. However, with the coupling coefficient $\chi = 3$ (see Fig.~\ref{ris:soliton_pm_chi}{\bf (e)}), the temporal profile of pulses deteriorates. In the case of the coupling coefficient $\chi = 1$, the solution found does not hold well in the time domain (see Fig.~\ref{ris:soliton_pm_chi}{\bf (f)}). Therefore, in the case of the coupling coefficient $\chi = 10$, the deformations of the MCF do not affect the found spatio-temporal soliton essentially. For a given coupling coefficient, the coupling length $1/\chi $ is ten times less than the dispersion length. Accordingly, the soliton manages to adapt itself every time to changes in the medium.

Results of numerical simulations demonstrate stability of the spatial distribution in the MCF within the considered parameter range ($\chi$, $\delta h/\chi$). The relative phase difference of the field in adjacent cores is $\pi$. As noted above, found solution \eqref{eq:6} is not destroyed in numerical simulation if the dispersion length is much less than the coupling length, i.e., $b^2 \ll \chi$. The black dotted lines in Fig.~\ref{ris:soliton_pm_chi}{\bf (g, h)} give an estimate of the fraction of the energy in the $\pm$-mode
\begin{equation}\label{eq:eta_pm}
\eta_\pm \approx 1-(\delta h/\chi)^2/2,
\end{equation}
which is the same as estimate \eqref{eq:pm_dP_lin}. 

Along with this, a series of numerical calculations was performed (with approximately 1000 realizations) when the refractive index changed randomly in the range $0 \leq \delta h_n \leq \delta h$. In this case, the $\pm$-mode was also preserved up to the coupling coefficient $\chi \geq 3$. In each numerical calculation, the fraction of the energy in the $\pm$-mode was random, but was limited to the region shown in Fig.~\ref{ris:soliton_pm_chi}{\bf (g, h)} with the yellow color. The blue dashed line shows the fraction of energy in the $\pm$-mode averaged over the ensemble. It follows from the figure that the average fraction of energy $\eta_\pm$ for random perturbations $\delta h_n$ is larger than that in the case of the statically bending MCF.

Thus, the found spatio-temporal soliton \eqref{eq:6} remains stable with respect to the MCF deformations in the case when the dispersion or nonlinear length is much greater than the coupling length, i.e., $\chi \gg b^2$. Stability estimate \eqref{eq:eta_pm} of the $\pm$-modes with respect to the deformation of the MCF structure ($\delta h_n \neq 0 $) and the amplification distribution over the MCF cores ($\gamma_n \neq 0$) can also be obtained analytically.

The appearance of large-scale deformations, as a rule, means that all linear modes in the MCF are excited, which makes the task difficult for analytical consideration. However, if the values $\delta h_n$ and $\delta \gamma_n = \gamma_n - \gamma$ are small (where $\gamma = \sum \gamma_n / 2N$), the perturbations of the $\pm$-mode will remain small. If the dispersion length is much less than the average gain length ($\gamma \ll B^2$), i.e., when the amplification of a soliton-like pulse occurs in the adiabatic mode, then an approximate solution of the equation \eqref{eq:3} for $N \ge 2$ can be sought in the form
\begin{equation}\label{eq:pm_ext}
		u_n=\sqrt{2} B \frac{(-1)^n+\delta_n}{\cosh(B \tau)} \rme^{2\rmi\chi z-\rmi B^2/2\gamma}, \quad B = b_0 \rme^{2\gamma z} . 
\end{equation}
Substituting this into Eq.~\eqref{eq:3}, we find in the first order of smallness with respect to $\delta_n, \delta h_n/\chi, \delta \gamma_n/\chi \ll 1$:
\begin{multline}\label{eq:pm_delta_eq}
\rmi \partial_z \delta_n \approx \chi (\delta_{n+1} + 2\delta_n + \delta_{n-1}) +\\+ (\delta h_n+\rmi \delta \gamma_n) (-1)^n.
\end{multline}
Here, we neglected the term ${4 B^2 \delta_n}/{\cosh^2(B \tau)}$ due to smallness of the coupling length compared to the dispersion length $B^2/\chi = b_0^2 \rme^{2\gamma z}/\chi \ll 1$ and the stability of solution \eqref{eq:6} with respect to small wave field perturbations (see Sec.~\ref{sec:4}). Note that the stability of solution \eqref{eq:6} to small field perturbations is also preserved in the active medium $\gamma \ne 0$ in the case $\gamma \ll B^2$. Equations~\eqref{eq:pm_delta_eq} are a system of coupled linear oscillators, which is driven by the force $(\delta h_n+\rmi \delta \gamma_n) (-1)^n$. Its forced (partial inhomogeneous) solution is easily found using the Fourier series expansion
\begin{multline}\label{eq:pm_ext_sol}
\delta_n \approx - \frac{1}{2 N} \sum_{m} \sum_{k\ne N} (\delta h_m + \rmi \delta \gamma_m) \dfrac{(-1)^m \rme^{\rmi \kappa_k (n-m)}}{4 \chi \cos^2 \frac{\kappa_k}{2}}.
\end{multline}

Typical examples of changes in the refractive index and the gain are
\begin{equation}\label{eq:dh_lin}
\delta h_n = \delta h \sin \frac{\pi n}{N}, \quad \delta \gamma_n = \delta \gamma \sin \frac{\pi n}{N},
\end{equation}
resulting from bending of optical fibers. Here, $\delta h$ and $\delta \gamma$ are the deformation amplitudes. For such disturbances, the form of the solution is much simpler:
\begin{equation}\label{eq:pm_lin_sol}
\delta_n \approx -\dfrac{\delta h+\rmi \delta \gamma }{4 \chi \sin^2\frac{\pi}{2N}} (-1)^n\sin\frac{\pi n}N.
\end{equation}
It can be seen from the obtained formulas that the perturbations remain small ($|\delta_n | \ll 1$) in the following cases:
\begin{itemize}
	\item the refractive index perturbation is small as compared to the coupling coefficient,
	\begin{equation}\label{eq:conditions_1}
			\delta h_n \ll 4\chi\sin^2\frac{\pi}{2N} ; 
	\end{equation}
	\item the average gain $\gamma$ is small as compared to the dispersion, and the coupling length is small as compared to the dispersion one,
	\begin{equation}\label{eq:conditions_2}
		\gamma \ll B^2 \ll \chi ;
	\end{equation}
	\item gain perturbations are small as compared to the coupling coefficient,
	\begin{equation}\label{eq:conditions_3}
			\delta\gamma \ll 4\chi\sin^2 \frac{\pi}{2N}.
	\end{equation}
\end{itemize}

Note that the fraction of the energy in the perturbed part is a quantity of the second order of smallness due to the orthogonality of the perturbations and the $\pm$-mode:
\begin{equation}\label{eq:pm_dP}
\delta W = \sum \int \frac{|u_n|^2 - 4 N B^2}{\cosh^2(B\tau)} d\tau \approx 4B \sum |\delta_n|^2 .
\end{equation}
In the particular case of perturbations \eqref{eq:dh_lin}, the expression for the fraction of energy in the perturbations takes a simple form
\begin{equation}\label{eq:pm_dP_lin}
\frac{\delta W}{W} \equiv 1 - \eta_\pm \approx
\frac{(\delta h/\chi)^2+(\delta \gamma/\chi)^2}{32 \sin^4 \frac{\pi}{2N}}.
\end{equation}
This estimate is in good agreement with the results of numerical simulation \eqref{eq:eta_pm} at $N = 3$ (black dotted lines in Fig.~\ref{ris:soliton_pm_chi}{\bf (g, h)}).

\section{Self-compression of $\pm$-solitons in active MCF}\label{sec:5}

This section presents studies of self-compression of out-of-phase soliton-like laser pulses propagating in an active MCF ($G(\omega) \neq 0$) with a finite gain bandwidth under the influence of Raman nonlinearity ($\mu \neq 0$). As noted above, the found stable nonlinear solution \eqref{eq:6} is determined by the $\pm$-mode in the transverse direction and has the form of an NSE soliton in the longitudinal direction it.

To obtain analytical estimates, we use the variational approach, which is generalized to the case of description of the nonlinear propagation of wave packets in non-conservative systems with a Gaussian form of the gain profile $G(\omega)$ on the frequency \eqref{eq:4}. This will reduce the partial differential equation to a closed system of ordinary differential equations for the characteristic parameters of a solitary laser pulse having a Gaussian form
\begin{equation}\label{eq:4_1}
u_n= (-1)^n \sqrt{\dfrac{W/2N}{\sqrt{\pi}\tau_p}}\rme^{-\frac{(\tau-q)^2}{2\tau_p^2}+\rmi\rho(\tau-q)^2-i\varpi(\tau-q)+\rmi\theta }.
\end{equation}
Here, parameters $\tau_p$, $\rho$, $\varpi$, $\theta$, $q$ characterize the pulse duration, chirp, frequency, and phase of the wave field at the center of intensity of the packet, whose position is determined by $q(z)$.

Despite the complication due to the absence of hamiltonianity in the active medium, the variational approach can be formulated for Eq.~\eqref{eq:3} too. At this, along with the Lagrangian of the conservative part,
\begin{multline}\label{eq:4_2}
\mathcal{L}=\sum_{n=1}^{2N} \int\limits_{-\infty}^{+\infty} \Big[ \dfrac{\rmi}2\left(u_n\dfrac{\partial u_n^*}{\partial z}-u_n^*\dfrac{\partial u_n}{\partial z}\right) - \dfrac12|u_n|^4 + \left|\dfrac{\partial u_n}{\partial\tau}\right|^2 - \\ 
- \chi (u_n u_{n+1}^* + u_{n+1} u_n^*)\Big]d\tau,
\end{multline}
it is necessary to determine the dissipative function
\begin{multline}\label{eq:4_7}
\delta\mathcal{Q}= \sum_{n=1}^{2N} \int\limits_{-\infty}^{+\infty} \dfrac{\rmi G(\omega)}{2\pi}\left[u_n(\omega)\delta u_n(\omega)^*-u_n^*(\omega)\delta u_n(\omega)\right]d\omega+\\
+\mu \sum_{n=1}^{2N} \int\limits_{-\infty}^{+\infty} \dfrac{\partial|u_n|^2}{\partial\tau} \delta|u_n|^2d\tau .
\end{multline}

Changing the parameters $a_j=\{W, \tau_p, q, \rho, \varpi,\theta\}$ of the distribution \eqref{eq:4_1} during the propagation of laser pulses is described by the Euler equations ($\dot{a}_j={da_j}/{dz}$) \cite{Ankiewicz07}
\begin{equation}\label{eq:4_5}
\dfrac{d}{dz}\dfrac{\partial\overline{\mathcal L} }{\partial \dot{a}_j}-\dfrac{\partial\overline{\mathcal L} }{\partial a_j}=\int \left[\frac{\delta\mathcal{Q}}{\delta u_n} \frac{\partial u_n}{\partial a_j} + \text{c.c.} \right] d\tau,
\end{equation}
where $\overline{\mathcal L}$ are the Lagrange functions \eqref{eq:4_2} calculated on the given distribution of the field \eqref{eq:4_1}.

Taking the integrals on the right-hand side of \eqref{eq:4_5}, we arrive at the following system of ordinary differential equations for the parameters of the wave field \eqref{eq:4_1}
\begin{subequations}\label{eq:4_9}
	\begin{gather}
	\dfrac{dW}{dz}=\dfrac{2W\tau_p\Omega\gamma}{\sigma} \rme^{-\tau_p^2\varpi^2/\sigma^2} , \label{eq:4_9a} \\
	\dfrac{d\tau_p}{dz}=4\rho\tau_p-\dfrac{\tau_p^2\Omega\gamma}{\sigma^5}\left(\dfrac{}{}4\tau_p^6\rho^2\Omega^2-\tau_p^2\Omega^2-8\tau_p^6\rho^2\varpi^2\right. +\nonumber \\ \left.\dfrac{}{} +2\tau_p^2\varpi^2+16\tau_p^8\rho^4-1 \right) \rme^{-\tau_p^2\varpi^2/\sigma^2} , \label{eq:4_9c} \\ 
	\dfrac{d\rho}{dz}=\dfrac1{\tau_p^4}-4\rho^2-\dfrac{W/2N}{\sqrt{8\pi}\tau_p^3}-\hspace{4cm}\nonumber \\-\dfrac{4\gamma\tau_p\rho\Omega}{\sigma^5}\left(\sigma^2-2\tau_p^2\varpi^2\right) \rme^{-\tau_p^2\varpi^2/\sigma^2} , \label{eq:4_9b} \\
	\dfrac{d\varpi}{dz}=-\dfrac{\mu W/2N}{\sqrt{2\pi}\tau_p^3}-\dfrac{2\tau_p\varpi\Omega\gamma}{\sigma^3}\left(1+4\tau_p^4\rho^2 \right) \rme^{-\tau_p^2\varpi^2/\sigma^2}, \label{eq:4_9d} \\
	\dfrac{dq}{dz}=\dfrac{4\tau_p^5\rho\varpi\Omega\gamma}{\sigma^3} \rme^{-\tau_p^2\varpi^2/\sigma^2} - 2\varpi , \label{eq:4_9e}
	\end{gather}
\end{subequations}
where $\sigma=\sqrt{\tau_p^2\Omega^2+4\tau_p^4\rho^2+1}$. Equation \eqref{eq:4_9e} for the velocity of the center of intensity of the wave packet is isolated from the rest of Eqs.~\eqref{eq:4_9}.

System of equations \eqref{eq:4_9} is greatly simplified if injected laser pulses have a soliton-like form corresponding to the equilibrium state of Eqs.~\eqref{eq:4_9c} and \eqref{eq:4_9b}. This gives a relation between the energy $W$, the frequency modulation $\rho$, and the laser pulse duration $\tau_p$:
\begin{equation}\label{eq:4_10}
W \approx 2N \dfrac{\sqrt{8\pi}}{\tau_p}, \quad \rho\simeq-\dfrac{\gamma}{4\tau_p^2\Omega^2}.
\end{equation}

\begin{figure}[t]
	\centering\includegraphics[width = \linewidth]{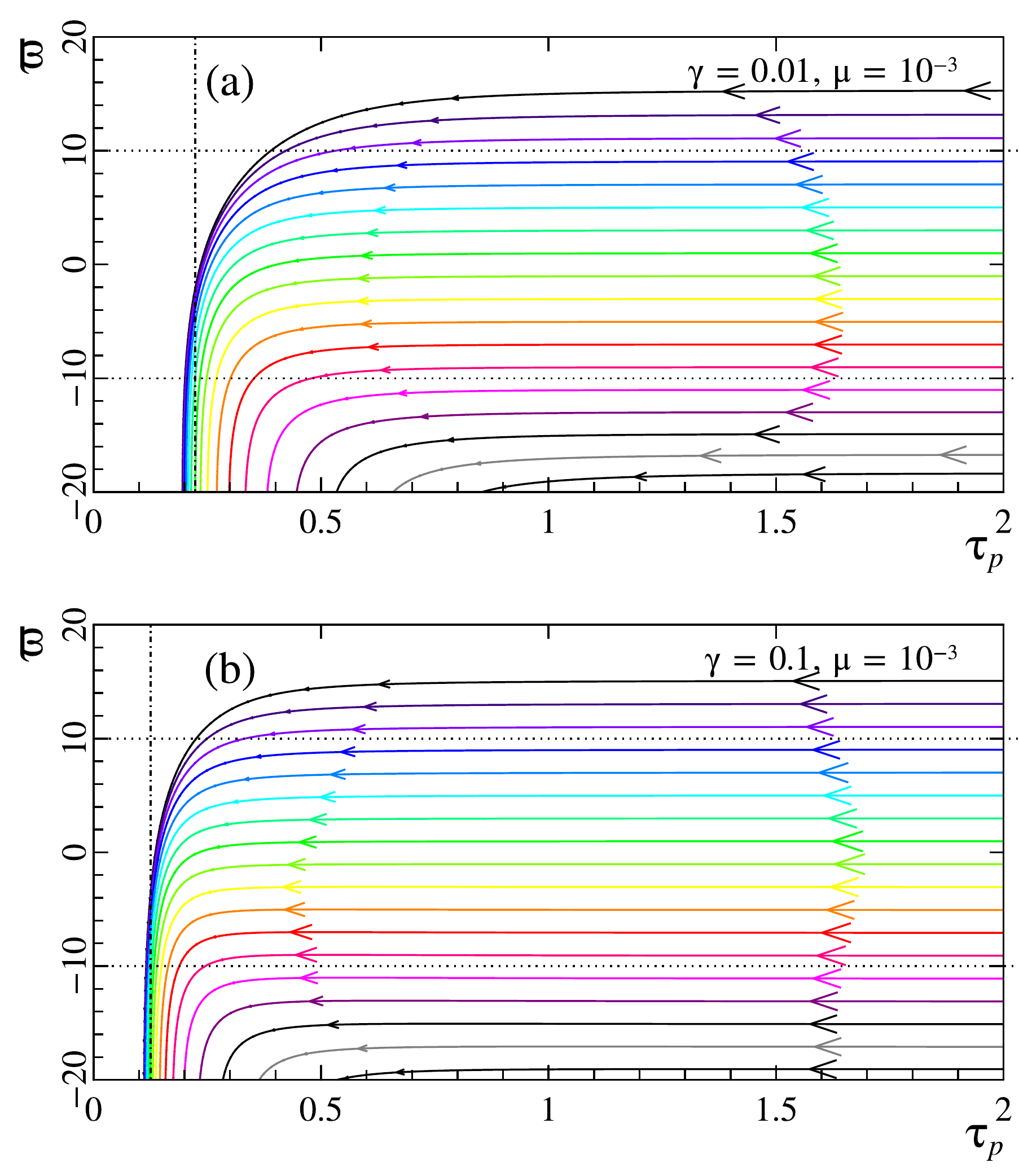}
	\caption{(Color online) The phase plane ($\tau_p,~\varpi$) of Eqs.~\eqref{eq:4_12} drawn for different values of the coefficients: {\bf (a)} $\gamma = 0.01$ , $\mu = 10^{-3}$; {\bf (b)} $\gamma = 0.1$, $\mu = 10^{-3}$. The vertical dashed line shows the estimate of the minimum laser pulse duration \eqref{eq:37}. The calculations were performed at $\Omega = 10$.
	}\label{ris:fig_ph_plain}
\end{figure}

We assume that the frequency modulation is small on the scale of the wave packet $\rho \tau_p^2 \ll 1$. As a result, we obtain the following system of equations that determines the decrease in the duration $\tau_p $ and the shift of the center frequency $\varpi$ down the spectrum of the wave packet of the soliton form
\begin{subequations}\label{eq:4_12}
	\begin{gather}
	\dfrac{d\tau_p}{dz}=-2\dfrac{\tau_p^2\Omega\gamma}{\sqrt{1+\tau_p^2\Omega^2}} \rme^{-\frac{\tau_p^2\varpi^2}{({1+\tau_p^2\Omega^2})^2}} , \label{eq:4_12a} \\
	\dfrac{d\varpi}{dz}=-\dfrac{2\mu}{\tau_p^4}-\dfrac{2\varpi\Omega\gamma\tau_p}{({1+\tau_p^2\Omega^2})^{3/2}} \rme^{-\frac{\tau_p^2\varpi^2}{({1+\tau_p^2\Omega^2})^2}} . \label{eq:4_12b}
	\end{gather}
\end{subequations}
The phase planes of this system of equations are shown in Fig.~\ref{ris:fig_ph_plain}.

The nonstationarity of the nonlinear response leads to the appearance of the minimal duration $\tau_\text{lim}$, to which the wave packet can be shortened during amplification. It follows from the Figure that at the initial stage, when the Raman nonlinearity does not affect the wave field dynamics, an adiabatic decrease in the wave packet duration at a constant center frequency is realized. At the final stage, the Raman response stops the decrease in the duration of the laser pulse due to a significant shift of the wave field spectrum to the long-wavelength region and to further removal of the radiation spectrum from the gain band of the active medium $\varpi \gtrsim \Omega$. It can be seen (Fig.~\ref{ris:fig_ph_plain}) that with a decrease in the gain $\gamma $ at a constant value of the coefficient $\mu$, an increase in the minimum value of the duration $\tau_\text{lim}$ of the compressed laser pulse takes place.

Let us estimate the minimal duration $\tau_\text{lim}$ depending on the parameters basing on Eqs.~\eqref{eq:4_12}. Let a laser pulse with the initial parameters be injected at the input of a nonlinear medium: $\tau_\text{in} \Omega \gg 1$ and $\varpi_\text{in} = 0$. As noted above, there is a decrease in the duration of the laser pulse at a practically constant frequency at the initial stage. So, Eq.~\eqref{eq:4_12a} reduces to
\begin{equation}\label{eq:4_13}
\dfrac{d\tau_p}{dz}\simeq-2\gamma\tau_p 
\end{equation}
and has the solution
\begin{equation}\label{eq:4_14}
\tau_p(z)=\tau_\text{in}\exp(-2\gamma z) .
\end{equation}
At the final stage, the frequency of the wave packet shifts down the spectrum rapidly due to the non-stationary nonlinear response of the medium, in accordance with Eq.~\eqref{eq:4_12b}
\begin{equation}\label{eq:4_15}
\dfrac{d\varpi}{dz} \simeq -\dfrac{2\mu}{\tau_p^4} .
\end{equation}
Substituting the expression for the wave packet duration \eqref{eq:4_14} into Eq.~\eqref{eq:4_15}, we find a solution for changing the carrier frequency:
\begin{equation}\label{eq:4_16}
\varpi(z)=-\dfrac{\mu}{4\gamma\tau_\text{in}^4}\exp(8\gamma z) \equiv - \frac{\mu}{4\gamma \tau_p^4}.
\end{equation}
The exponential decrease in the laser pulse duration will stop when the central radiation frequency is shifted beyond the medium gain band $|\varpi| \sim \Omega$. This gives an estimate of the minimal duration of the compressed laser pulse depending on the media parameters:
\begin{equation}\label{eq:37}
\tau_\text{lim}=\left(\dfrac{\mu}{4\gamma\Omega}\right)^{1/4} .
\end{equation}
It can be seen from this expression that large gain values $\gamma$ yield shorter wave packet durations.

In Figure \ref{ris:fig_ph_plain}, the black dotted line shows the position of the minimal duration $\tau_\text{lim}$. The estimated value of $\tau_\text{lim}$ is in good agreement with the results of a numerical analysis of the phase plane of Eqs.~\eqref{eq:4_9}. Note that in the case when the carrier frequency of the injected wave packet is additionally shifted up the spectrum $\varpi_\text{in} = \Omega$, the degree of compression of the laser pulse can be slightly increased by approximately $\sqrt[4]{2} \approx 1.19$ times.

In dimensional variables, the compression length, the minimum duration, and the maximum laser pulse energy are
\begin{subequations}\label{eq:lim}
\begin{gather}
	L_\text{comp} =\frac1{2\gamma}\ln\frac{\tau_\text{in}}{\tau_\text{lim}} =\frac1{2\gamma}\ln\left[\tau_\text{in}\sqrt[4]{\frac{8\gamma\Omega}{\tau_R\beta_2}} \right] , \label{eq:L_comp} \\
	\tau_\text{lim}=\sqrt[4]{\frac{\tau_R\beta_2}{8\gamma\Omega}} , \label{eq:tau_lim} \\ 
	W_\text{lim}=\frac{4\sqrt{\pi}N}{\Gamma}\sqrt[4]{\frac{2 \beta_2^3\Omega\gamma}{\tau_R}} . \label{eq:W_lim}
\end{gather}
\end{subequations}
Note, that the wave packet duration at half maximum intensity is $\tau_\text{lim}^\text{\tiny FWHM}=2 \sqrt{\ln2}\tau_\text{lim}$.

Recall that the proposed method for decreasing the duration of the out-of-phase soliton-like laser pulse in the active MCF adiabatically is possible under applicability conditions \eqref{eq:conditions_1}, \eqref{eq:conditions_2}, and \eqref{eq:conditions_3}.

\section{Numerical simulations}\label{sec:6}

To illustrate the proposed method of laser pulse self-compression, we turn to the results of numerical simulation within the framework of initial equation \eqref{eq:2} with the nonlinear response $P_n^\text{NL}$ \eqref{eq:2_1}. A laser pulse at the wavelength $\lambda = 2~\mu$m with an initial duration of $\tau^\text{\tiny FWHM}_\text{in}=1$~ps (half-height intensity) was injected into the input of a 6-core silica MCF ($N = 3$) doped with thulium. The calculations were performed for a fiber with the following parameters: the group velocity dispersion $\beta = 100$~ps$^2$/km, the nonlinearity coefficient $\Gamma = 1$/(W\,km), and the gain band $1/\Omega = 30$~fs \cite{Jackson}.

\begin{figure*}
	\includegraphics[width = 0.8\linewidth]{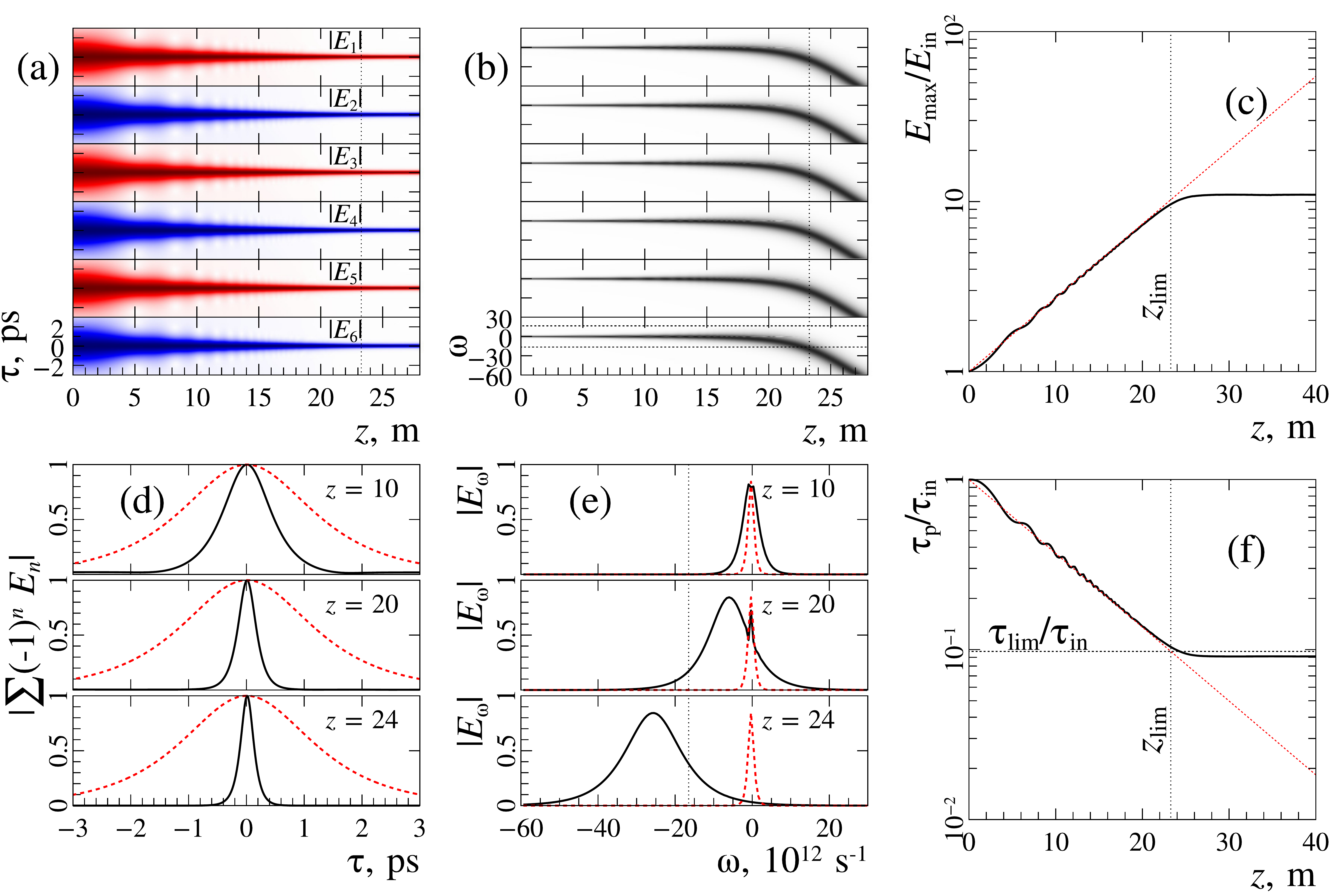}
	\caption{(Color online) Dynamics of the wave packet ${E}_n(z,\tau)$ {\bf (a)} and its spectrum ${E}_n(z,\omega)$ {\bf (b)} along the propagation path $z$ (the electric field strength and its spectrum are normalized to their maximal values), the maximal amplitude {\bf (c)}, and the laser pulse duration {\bf (f)} depending on $z$. Figures {\bf (d)} and {\bf (e)} show the distribution of the total wave field $|\sum (-1)^n {E}_n(\tau)|$ and its spectrum for different $z$. The horizontal dashed line in Fig.~{\bf (f)} shows the value \eqref{eq:tau_lim}. The vertical dashed line in Fig.~{\bf (e)} is the boundary of the gain band of the active medium $\omega=-\Omega/2=-17\times10^{12}~\text{s}^{-1}$. The vertical dashed lines in Figs.~{\bf (a, b, c, f)} are the media length $L_\text{comp} \approx 23$~m, at which the wave packet duration reaches the minimal value \eqref{eq:L_comp} $\tau_\text{lim}^\text{FWHM}\approx 97$~fs. The calculations were performed for $\tau_\text{in}^\text{\tiny FWHM}=1$ ps, $\beta=100~\text{ps}^2/\text{km}$, $1/\gamma=20$~m, $1/\Omega=30$~fs, $\chi^{-1}=3$~mm, $\Gamma=1~\text{(W km)}^{-1}$.
	}\label{ris:fig_ampl_G01}
\end{figure*}

Figure \ref{ris:fig_ampl_G01} shows the evolution of a laser pulse in a medium with the gain $\gamma^{-1} = 20$~m, which corresponds to the adiabatic mode of wave packet amplification (the dispersion length is less than the gain length). A laser pulse was injected at the MCF input in the form of the found spatio-temporal soliton \eqref{eq:6}
\begin{equation}
	{E}_n=(-1)^n\frac{17.54}{\cosh(\tau/\tau_\text{sol}^\text{in})},
\end{equation}
where $|{E}_n|^2$ is measured in watts, $\tau_\text{sol}^\text{in} = \frac{1}{ 2\mathop{\mathrm{acosh}}\sqrt{2} } \tau^\text{\tiny FWHM}_\text{in}=0.57$~ps is the duration of the soliton.  The dispersion length is $(\tau_\text{sol}^\text{in})^2/2 \beta \approx 1.6$~m. This corresponds to the initial energy in the MCF with six cores, $W^\text{6 cores}_\text{in}=2.1$~nJ. From expressions \eqref{eq:lim} it follows that at the MCF output $L_\text{comp}\approx 23$~m long, the wave packet duration will reach the minimum value $\tau_\text{lim}^\text{\tiny FWHM} \approx 97$~fs. To ensure the regime of stable laser pulse propagation in the MCF, it is necessary to choose the coupling coefficient $\chi$ so that the coupling length is much less than the dispersion length \eqref{eq:conditions_2} estimated for the minimal duration $\tau_\text{lim}^\text{\tiny FWHM}$. This corresponds to the coupling coefficient $\chi^{-1} < 0.5$~cm.

Figure \ref{ris:fig_ampl_G01}{\bf (a)} shows the dynamics of the wave packet $u_n (z, \tau)$ and its spectrum \ref{ris:fig_ampl_G01}{\bf (b)} along the propagation path $z$. The electric field strength and its spectrum are normalized to their maximum values. The horizontal dashed lines in Fig.~\ref{ris:fig_ampl_G01}{\bf (b)} show the boundary of the gain band $\omega = -\Omega/2 = -17 \times 10^{12}~\text{s}^{-1}$. As can be seen from Fig.~\ref{ris:fig_ampl_G01}{\bf (a)}, the $\pm$-mode is preserved during the laser pulse evolution in the active MCF (the relative phase difference of the fields between neighboring nuclei is $\pi$). Figures~\ref{ris:fig_ampl_G01}{\bf (d)} and \ref{ris:fig_ampl_G01}{\bf (e)} show the distribution of the total wave field $|\sum (-1)^n {E}_n(\tau)|$ and its spectrum distribution for different $z$. The red dotted line shows the initial distribution, the black line shows the current distribution, and the dashed vertical line shows the border of the gain band. At the initial stage ($z \lesssim 20$~m), as the wave packet propagates in the medium, an adiabatic decrease in the laser pulse duration takes place (see Fig. \ref{ris:fig_ampl_G01}{\bf (a)}). At this, the spectrum broadens uniformly in both directions at a constant center frequency of the laser pulse (see Figs.~\ref{ris:fig_ampl_G01}{\bf (b)} and \ref{ris:fig_ampl_G01}{\bf (e)} at $ z = 10$~m). Subsequently, the spectrum of the wave field begins to shift to the long-wave part due to the influence of non-stationary non-linear response. The laser pulse spectrum goes beyond the gain band at $z \approx 24$~m (see Fig.~\ref{ris:fig_ampl_G01}{\bf (b, e)}), and the decrease in the duration of the wave packet stops. Further, the center of the laser pulse spectrum monotonically shifts to the long-wavelength region for a constant duration of the wave packet (see Fig.~\ref{ris:fig_ampl_G01}{\bf (b)}).

\begin{figure*}
	\includegraphics[width = 0.8\linewidth]{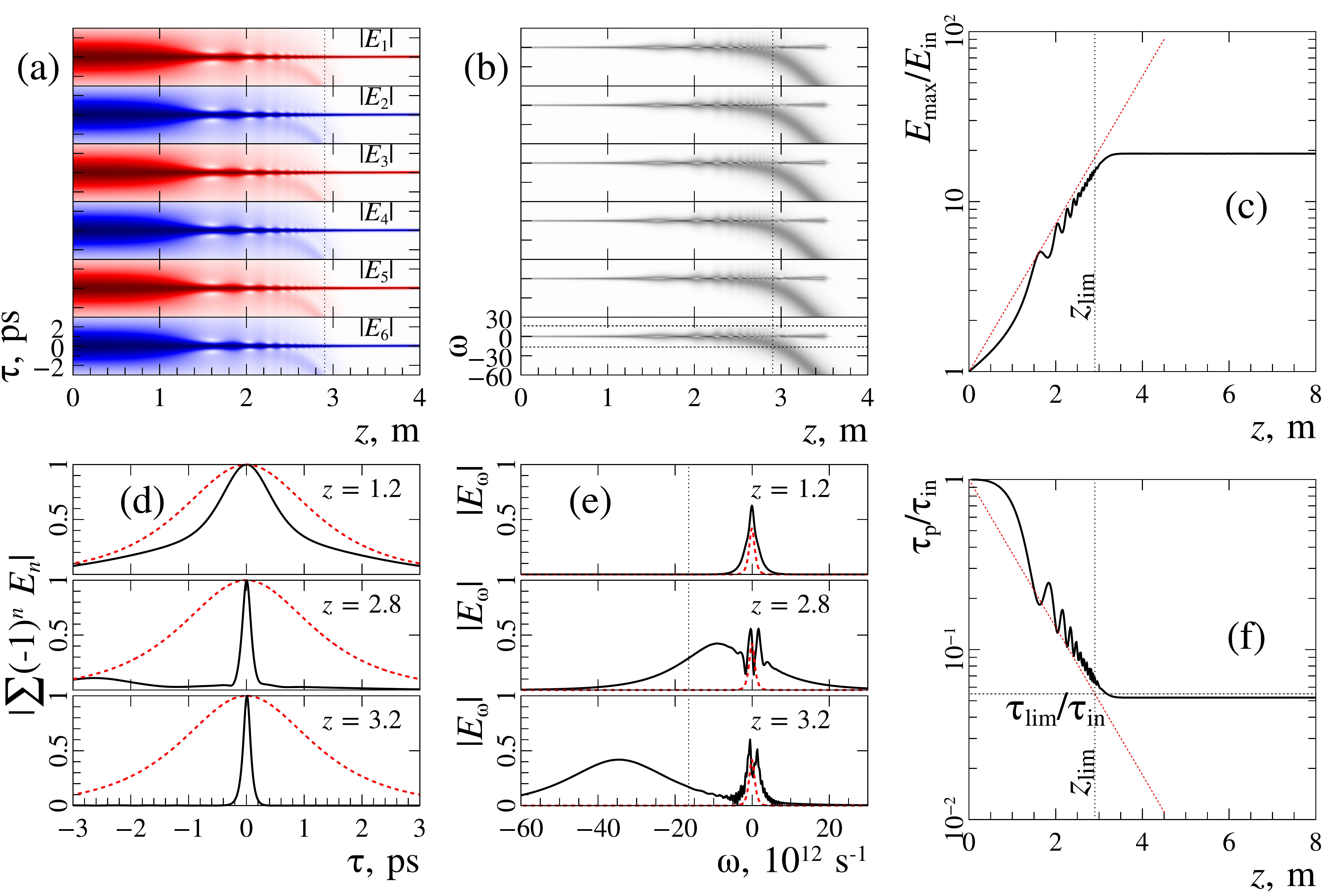}
	\caption{(Color online) Same as in Fig.~\ref{ris:fig_ampl_G01}. The calculations were performed for $\tau_\text{in}^\text{FWHM}=1$ ps, $\beta=100~\text{ps}^2/\text{km}$, $1/\gamma=2$ m, $1/\Omega=30$ fs, $\chi^{-1}=3$~mm, $\Gamma=1~\text{(W km)}^{-1}$. The vertical dashed line in Fig. {\bf (a, b, c, f)} is the medium length $L_\text{comp} \approx 2.9$ m, at which the wave packet duration reaches the minimal value \eqref{eq:L_comp} $\tau_\text{lim}^\text{FWHM}\approx 55$~fs.
	}\label{ris:fig_ampl_G1}
\end{figure*}

Figures \ref{ris:fig_ampl_G01}{\bf (c,f)} show the dependences of the maximum amplitude ${E}_\text{max}$ {\bf (c)} and the wave packet duration $\tau_p$ {\bf (d)} along the propagation path of $z$. Values are normalized to their initial values. 
The red dotted line shows the approximation of the laser pulse parameters $u_\text{max}\propto \rme^{2\gamma z}$, $\tau_p\propto\rme^{-2\gamma z}$. 
It can be seen that throughout the evolution of the wave packet in the MCF, the adiabatic decrease in the duration of the found spatio-temporal soliton \eqref{eq:6} is maintained. The vertical line in Fig.~\ref{ris:fig_ampl_G01}{\bf (c,f)} shows the position of the estimated compression length $L_\text{comp}$ \eqref {eq:L_comp}, and the horizontal dashed line in Fig.~\ref{ris:fig_ampl_G01}{\bf (f)} shows the position of the minimal duration $\tau_\text{lim}$ \eqref{eq:tau_lim}. It is seen that the maximum increase in the field amplitude and the maximum decrease in the duration of the laser pulse are achieved on the path $z\approx L_\text{comp}$. Subsequently, the considered quantities reach stationary values. Therefore, the obtained estimates of the minimal duration of the wave packet $\tau_\text{lim}$ and the compression length $L_\text{comp}$ are in good agreement with the results of numerical simulation.

Thus, at the output of an active MCF of length $L_\text{comp}\approx 23$~m, the wave packet is shortened by about 10 times from $\tau_\text{in}^\text{\tiny FWHM} = 1$~ps to $\tau_\text{out}^\text{\tiny FWHM} \approx 97$~fs for the gain $\gamma^{-1} = 20$~m. However, implementation of such an extended active medium seems difficult. In this regard, we consider an MCF with a larger gain $\gamma^{-1} = 2$~m. An increase in the value of the parameter $\gamma$ by an order of magnitude should lead to a decrease in the compression length and to a decrease in the minimal wave packet duration in comparison with the case $\gamma^{-1} = 20$~m, according with estimate \eqref{eq:lim}.

Figure \ref{ris:fig_ampl_G1} shows the dynamics of the wave packet and its spectrum. Three stages of the evolution of a laser pulse can be distinguished here. The first stage is characterized by a noticeable discharge of radiation in the time domain (Fig.~\ref{ris:fig_ampl_G1}{\bf (a)} at $z \lesssim 1.8$~m). The reason is a violation of the adiabatic amplification condition for a soliton-shaped pulse (when the dispersion length is small with the gain length) for such a high value of the parameter $\gamma$. At this stage, the dependence of the maximum amplitude and minimum duration on the evolution variable $z$ differs significantly from the exponential law $u_\text{max} \propto 1/\tau_p \propto \rme^{2 \gamma z}$ (Fig.~\ref{ris:fig_ampl_G1}{\bf (c,f)}). A decrease in the laser pulse duration by several times leads to the second stage, when the pulse duration and maximum amplitude begin to change exponentially (quasi-adiabatic regime), since at this stage the dispersion length becomes less than the gain one. At this stage, there is no radiation discharge. In accordance with the estimate \eqref{eq:lim}, the wave packet duration reach the minimal value of $\tau_\text{lim}^\text{\tiny FWHM} \approx 55$~fs at the MCF output $L_\text{comp} \approx 2.9$~m. Despite the initially non-adiabatic regime of amplification, the obtained estimates of the minimal laser pulse duration $\tau_\text{lim}$ and the compression length $L_\text{comp}$ are in good agreement with the results of numerical simulation (Fig.~\ref{ris:fig_ampl_G1}{\bf (c,f)}). At the third stage ($z \gtrsim 3.2$~m), the duration of the laser pulse does not change, since the spectrum of the laser pulse has shifted beyond the gain band.

It should be noted that due to the non-stationarity of the nonlinear response, the main signal is purified from the background radiation generated in the first stage due to the difference in group velocities. This is clearly seen in Fig.~\ref{ris:fig_ampl_G1}{\bf (d)} at $z = 2.8$~m (the background radiation is located to the left of the main signal). Moreover, this difference in speeds increases with the growth of the $z$ path, since the frequency of the main signal decreases as it propagates in media. As a result, the spatio-temporal soliton is purified of the background radiation quickly.

So, the performed qualitative analysis based on the variational approach is in good quantitative agreement with the results of numerical simulations performed within the framework of initial equation \eqref{eq:2}.

So, at the output of a three-meter long MCF, it is possible to shorten the laser pulse from $\tau_\text{in}^\text{\tiny FWHM}=1$~ps to $\tau_\text{lim}^\text{FWHM}\approx 55$~fs (Fig.~\ref{ris:fig_ampl_G1}{\bf (d)}). In this case, the energy contained in the space-time soliton at the output of a 6-core MCF is $W^\text{6 cores}_\text{lim} \approx 38.4$~nJ (energy in one core is $W_\text{lim}^\text{1 core} \approx 6.4$~nJ for these parameters). This energy can be significantly increased if one uses a large number of cores on the ring $N \ggg 1$. For example, in the case of 240 cores ($N = 120$), the energy in the laser pulse is $W_\text{lim}^\text{240 cores} = 1.5$~$\mu$J. Unfortunately, the compression in MCF from such a large number of cores becomes sensitive to MCF deformations \eqref{eq:conditions_1} and the gain inhomogeneities \eqref{eq:conditions_3}, which contributes to the rescattering of the $\pm$-modes to other modes (see section \ref{sec:3}). As a result, the maximum achievable energy of the output wave packet in the proposed scheme is determined only by the technological capabilities of manufacturing MCF with identical cores (see table~\ref{tbl:1}). Gain homogeneity requirements $\tfrac{\delta \gamma}{\gamma} \ll \tfrac{2\chi}{\gamma} \tfrac{\pi^2}{N^2} \le 1$ are significantly weaker due to the large factor $2 \chi/\gamma \sim 10^3$ for typical parameters $\chi^{-1} < 0.5$~cm, $\gamma^{-1} = 2$~m.

\begin{table}[tb]
	\caption{Maximal energy and requirements for $\delta h$ for a different number of cores in the MCF.} \label{tbl:1}
	\begin{tabular}{|c|c|c|}
		\hline No. of cores ($2N$)& $W_\text{lim}$ & $\delta h/\chi$ \\\hline
		6 & $\approx 38.4$ nJ & $\le 1$ \\\hline
		20 & $\approx 128$ nJ & $\le 0.1$ \\\hline
		60 & $\approx 384$ nJ & $\le 0.01$ \\\hline
		200 & \quad$\approx 1.28$ $\mu$J~~~ & \quad $\le 0.001$~~~ \\\hline
	\end{tabular}
\end{table}

\section{Conclusion}

In this work we consider the propagation of out-of-phase wave packets in a multi-core fiber (MCF) made of an even number of cores located in a ring. The exact nonlinear solution was found \eqref{eq:6} in the form of solitons with out-of-phase distribution in the transverse direction. Its stability is proved with respect to both small perturbations of the wave field, including azimuthal ones, and small deformations of the MCF structure. The proof of existence of a stable out-of-phase solitons opens up wide possibilities for its use as an analog of the fundamental mode for the considered MCFs with the goal of significantly increasing the energy transported through the fiber.

As an example of using this soliton distribution, we studied the problem of compression of laser pulses in an active MCF with a sign-constant gain profile of a Gaussian form. The qualitative analysis of the wave packet self-action in the considered MCF is carried out basing on the variational approach. The optimal fiber parameters, the minimal duration of the output pulse, and the compression length are found that are in good agreement with the results of numerical simulation. The requirements for MCF deformations and for the gain magnitude and uniformity are determined in order to achieve high energies in the output laser pulse.

This work was supported by the RSF No. 16-12-10472.

\end{document}